\documentclass[aps,pra,onecolumn,amsmath,amssymb,latexsym,floatfix,,preprint,superscriptaddress,nofootinbib]{revtex4}
\usepackage{color,amssymb}
\usepackage[dvips]{graphicx}
\usepackage{amsmath,psfrag}
\usepackage{psfrag}
\usepackage{longtable}
\usepackage{dcolumn,epsfig}
\usepackage[normalem]{ulem}
\usepackage{color}

\newcommand{\comment}[1]{}

\newcommand{\be}{\begin{equation}}
\newcommand{\ee}{\end{equation}}
\newcommand{\ba}{\begin{eqnarray}}
\newcommand{\ea}{\end{eqnarray}}

\begin{document}
\title{ Quantum ground-state cooling and tripartite entanglement with three-mode optoacoustic interactions}
\author{H. Miao}
\author{C. Zhao}
\author{L. Ju}
\author{D. G. Blair}
\affiliation{School of Physics, University of Western Australia (UWA), WA 6009, Australia}

\begin{abstract}
We present a quantum analysis of three-mode optoacoustic parametric interactions in an optical
cavity, in which two orthogonal transverse optical-cavity modes are coupled to one acoustic mode
through radiation pressure. Due to the optimal frequency matching --- the frequency separation of
two cavity modes is equal to the acoustic-mode frequency --- the carrier and sideband fields simultaneously
resonate and coherently build up. This mechanism significantly enhances the optoacoustic couplings
in the quantum regime. It allows exploration of quantum behavior of optoacoustic interactions in
small-scale table-top experiments. We show explicitly that given an experimentally achievable
parameter, three-mode scheme can realize quantum ground-state cooling of milligram scale mechanical
oscillators and create robust stationary tripartite optoacoustic quantum entanglements.
\end{abstract}
\maketitle

\section{Introduction}
Optoacoustic interactions have recently become of great interest, for their potential in exploring
the quantum behavior of macroscopic objects. Various experiments have demonstrated that the acoustic
mode of a mechanical oscillator can be cooled significantly through two-mode optoacoustic interactions
\cite{Cohadon, Naik, Gigan, Arcizet, Kleckner,Schliesser1, Corbitt1, Corbitt2, Schliesser2, Poggio,
Thompson, Lowry, Schediwy}. The basic setup consists of a Fabry-Perot cavity with an end mirror. Linear
oscillations of the mirror acoustic mode at frequency $\omega_m$ scatter the optical-cavity mode
(usually $\rm{TEM}_{00}$) at frequency $\omega_0$  into Stokes ($\omega_0-\omega_m$) and anti-Stokes
($\omega_0+\omega_m$) sideband modes which have the same spatial mode shape as the $\rm{TEM}_{00}$ mode.
The optical cavity is appropriately detuned such that the anti-Stokes sideband is close to resonance.
Therefore, the anti-Stokes process is favored over the Stokes process. As a natural consequence of
energy conservation, the thermal energy of the acoustic mode has to decrease in order to create high-energy
anti-Stokes photons at $\omega_0+\omega_m$. If the cavity-mode decay rate (related to the optical finesse)
is smaller than the acoustic-mode frequency, theoretical analysis shows that these experiments can
eventually achieve the quantum ground state of a macroscopic mechanical oscillator \cite{Marquardt, Rae,
Genes}, which would be a significant breakthrough in physics from both experimental and theoretical
points of view. With the same scheme, many interesting issues have been raised in the literature, such
as teleportation of a quantum state into mechanical degrees of freedom \cite{Mancini2}, creation
of stationary quantum entanglements between the cavity mode and the mechanical oscillator \cite{Vitali,Paternostro}
or even between two oscillators \cite{Helge, Hartmann}. This in turn could be implemented in future
quantum communications and computing.

The concept of three-mode optoacoustic parametric interactions was first introduced and analyzed theoretically
in the pioneering work of Braginsky {\it et al.} \cite{braginsky}. It was shown that three-mode interactions inside
high-power optical cavities of large-scale laser interferometric gravitational-wave detectors have the
potential to induce instabilities, which would severely undermine the operation of detectors. This analysis
was elaborated by many other authors to more accurately simulate the real situation in next-generation advanced
gravitational-wave detectors \cite{braginskyII,zhao, Ju, Gurkovsky} and to find strategies for suppressing instability \cite{braginskyIII, Gras}.
Recently, the UWA group experimentally demonstrated three-mode interactions in an $80$-m high-power optical
cavity by exiting acoustic modes and observing resonant scattering of light into a transverse cavity mode
\cite{AIGO}.

Different from the two-mode case, in three mode interactions, a single acoustic mode of the mechanical
oscillator scatters the main cavity $\rm{TEM}_{00}$ mode into another transverse cavity mode which has a
different spatial distribution from the $\rm{TEM}_{00}$ mode. Specifically, when the $\rm{TEM}_{00}$ mode
is scattered by the acoustic mode, the frequency is split into Stokes and anti-Stokes sidebands at $\omega_0\pm\omega_m$,
and in addition, the spatial wave front is also modulated by the acoustic mode. Three-mode interactions happen strongly
when both the modulation frequency and spatial mode distribution are closely matched to those of another
transverse optical-cavity mode. Under these circumstances, both the carrier and sideband
modes are simultaneously resonant inside the cavity and get coherently build up. Taking into account the
resonance of the acoustic mode, the system is triply resonant, with the interaction strength scaled by
the product of the two optical quality factors and the acoustic quality factor. If the transverse optical
cavity mode has a frequency lower than the ${\rm TEM}_{00}$ mode, the Stokes sideband will be on resonance
and the interaction provides positive amplification of the acoustic mode, while if the transverse mode has
a frequency above the main cavity mode ${\rm TEM}_{00}$, the anti-Stokes sideband will resonate, and the
system has negative gain and the acoustic mode will be cooled. The underlying principle of both two and three mode
interactions is similar to the Brillouin scattering, except
that the modulation occurs not through changes in refractive index of the medium, but through bulk surface motion
of a macroscopic mechanical oscillator (i.e. the acoustic mode) which modulates the optical path of the light.

While three-mode interactions are inconvenient byproducts of the design of advanced gravitational-wave detectors,
they can be engineered to occur in small scale systems with low mass resonators, which can serve as an optoacoustic
amplifier and be applied to acoustic-mode cooling \cite{zhao2,Miao2}. Besides, due to its triply resonant feature, the
three-mode system has significant advantages
compared with the two-mode system and allows much more stronger optoacoustic couplings.
To motivate experimental realizations, we have suggested a small-scale table-top experiment with a milligram
mechanical oscillator in a coupled cavity \cite{Miao2}. Using the extra degree of freedom of the coupled cavity,
the cavity mode gap ( the difference between the two relevant cavity modes) can be continuously tuned such that
it is equal to plus or minus the acoustic-mode frequency, which maximizes the three-mode interactions strength. We
also pointed out that, in the negative-gain regime, this experimental setup can be applied to resolved-sideband
cooling of a mechanical oscillator down to its quantum ground state. In that paper, we used the classical
analysis presented by Braginsky {\it et al.} to obtain the effective thermal occupation number $\bar n$ of the
acoustic mode. This analysis {\it breaks down} when $\bar n\ll1$ and the quantum fluctuations of the cavity modes
have to be taken into account. To overcome this limitation, we used the similarity in the Hamiltonian of the two-mode
and the three-mode system, and argued that the quantum limit for cooling in both systems is the same without investigating
the detailed dynamics.  However, in order to gain a quantitative understanding of the three-mode
system in the quantum regime, it is essential to develop a full quantum analysis which includes the dynamical
effects of the quantum fluctuations. Besides, as we will show, the quantum analysis reveals a most interesting
non-classical feature of three-mode systems: stationary tripartite quantum entanglement.

The outline of this paper is the following: In Sec. \ref{II},  we  start from the classical analysis given by
Braginsky {\it et al.} and then quantize it with the standard approach. In Sec. \ref{III}, we use this quantized
Hamiltonian as the starting point to analyze the dynamics of the three-mode system. Further, based upon the
 Fluctuation-Dissipation-Theorem (FDT), we derive the quantum limit for the achievable thermal occupation
number in cooling experiments. To motivate future small-scale experiments, we provide an experimentally-achievable
specification for the quantum ground-state cooling of a mechanical oscillator. In Sec. \ref{IV}, we investigate
the stationary tripartite optoacoustic quantum entanglement and we show that the same specification for the cooling
experiments can also be applied to realize robust stationary optoacoustic entanglements.

\section{Quantization of three-mode parametric interactions \label{II}}

In this section, we will first present the classical formulations of three-mode optoacoustic parametric interactions
given by Braginsky {\it et al.} \cite{braginsky} and then apply standard procedures to obtain the quantized version.

\begin{figure}
\includegraphics[width=0.5\textwidth, bb=91 325 416 437,clip]{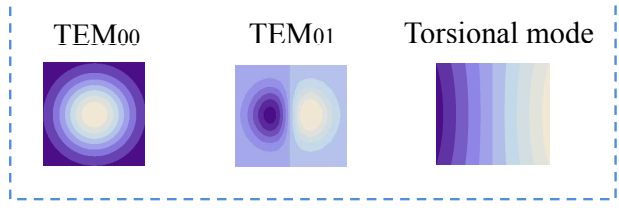}\caption{Spatial shapes of the $\rm TEM_{00}$ and $\rm TEM_{01}$ modes
and the acoustic torsional mode.}
\label{modes}
\end{figure}

\noindent{\it Classical Picture.} A detailed quantitative classical formulation of three-mode interactions was
given in the Appendix of Ref. \cite{braginsky}. A Lagrangian formalism was used to derive the classical equations
of motion and analyze the stability of the entire three-mode optoacoustic system. The formalism can be easily converted
into Hamiltonian language, which can then be quantized straightforwardly. For convenience, we will use slightly different
notation and definitions for the optical fields. Further, we assume that the two optical-cavity modes are the
${\rm TEM}_{00}$ and ${\rm TEM_{01}}$ modes and the acoustic mode has a torsional mode shape (about vertical axis) which has
a large spatial overlap with the $\rm TEM_{01}$ mode as shown in Fig. \ref{modes}. It can be easily extended to general
cases with other transverse optical modes and acoustic modes. Assuming the electric field is linearly polarized in the
transverse direction perpendicular to $z$ axis, the electromagnetic fields ($E, H$) of the cavity modes can be written as
\ba
E_{i}(t)&=& \left(\frac{\hbar\,\omega_i}{\epsilon_0 V}\right)^{1/2}f_i(\vec{r}_{\bot})\sin(k_i z) q_i(t),\\
H_{i}(t)&=& \frac{\epsilon_0}{k_i}\left(\frac{\hbar\,\omega_i}{\epsilon_0 V}\right)^{1/2} f_i(\vec{r}_{\bot})\cos(k_i z) \dot q_i(t).
\ea
Here $i=0,1$ represent the ${\rm TEM}_{00}$ and ${\rm TEM_{01}}$ modes; $f_i(\vec{r}_{\bot})$ are the transverse mode shapes;
$\omega_i$ denote the eigenfrequency; $k_i$ are the wave numbers; $V$ is the volume of the optical cavity; $q_i(t)$ are the
generalized coordinates of the fields; $\dot q_i(t)$ are the time derivatives of $q_i(t)$. At present stage, the appearance of $\hbar\,\omega_i$ is just to make the generalized coordinates $\hat q_i$ dimensionless.
The classical Hamiltonian of this system is given by
\be
{\cal H}={\cal H}_m+\frac{1}{2}\int d\vec{r}_{\bot} (L+x\,u_z)[\epsilon_0(E_0+E_1)^2+\mu_0(H_0+H_1)^2],
\ee
where $L$ is the length of the cavity; $x$ is the generalized coordinate of the acoustic mode; $u_z$ is the vertical displacement.
The free Hamiltonian of the acoustic mode is
\be
{\cal H}_m=\frac{1}{2}\hbar\,\omega_m (q_m^2+p_m^2)
\ee
with $q_m\equiv x/\sqrt{\hbar/(m\omega_m)}$ and $p_m$ is the momentum normalized with respect to $\sqrt{\hbar m\omega_m}$.

After integrating over the transverse direction and taking into account of the mode shapes, we obtain
\be
{\cal H}={\cal H}_m+{\cal H}_0+{\cal H}_1+{\cal H}_{\rm int}.
\ee
Defining dimensionless canonical momentum $p_i(t)\equiv \dot q_i(t)/\omega_i$, the free Hamiltonian of the two cavity modes are
\be
{\cal H}_i=\frac{1}{2}\hbar\,\omega_i(q_i^2+p_i^2)
\ee
and the interaction Hamiltonian is given by
\be
{\cal H}_{\rm int}=\hbar\, G_0 q_m(q_0q_1+p_0p_1),
\ee
where the coupling constant is defined as $G_0\equiv \sqrt{\Lambda\,\hbar\,\omega_0\omega_1/(m\,\omega_m L^2)}$ with the geometrical overlapping factor $\Lambda\equiv (L\int d\vec{r}_{\bot}u_z f_0 f_1/V)^2$.

Given above Hamiltonian, it is straightforward to derive the classical equations of motion and analyze the dynamics of the system, which
would be identical to those in the Appendix of Ref. \cite{braginsky}.
To quantify the strength of three-mode interactions,  Braginsky {\it et al.} introduced the parametric gain ${\cal R}$, as defined by
\be\label{R}
{\cal R}=\pm\frac{2\Lambda I_0 \omega_1}{m\,\omega_m L^2\gamma_0\gamma_1\gamma_m}=\pm \frac{2\Lambda I_0 Q_0Q_1Q_m}{m\,\omega_0\omega_m^2L^2}
\ee
where $\pm$ correspond to either positive gain or negative gain;
$I_0$ is the input optical power of the ${\rm TEM}_{00}$ mode; we have defined optical and
acoustic-mode quality factors $Q_i=\omega_i/\gamma_i\,(i=0,1,m)$. Due to optoacoustic interaction, the decay rate $\gamma_m$ of
the acoustic mode will be modified to an effective one $\gamma_m'$, which is
\be
\gamma_m'\approx (1-{\cal R})\gamma_m.
\ee
When ${\cal R}>1$, the decay rate becomes negative and this corresponds to instability. In this paper, we
are particularly interested in the regime where ${\cal R}<0$ which gives rise to the acoustic-mode cooling.
The effective thermal occupation number $\bar n_{\rm th}'$ of the acoustic mode is given by
\be\label{nb}
\bar n_{\rm th}'=\frac{\bar n_{\rm th}\gamma_m}{\gamma_m'}=\frac{\bar n_{\rm th}}{1-{\cal R}}
\ee
with $\bar n_{\rm th}$ denoting the original thermal occupation number. It looks as if $-{\cal R}\gg 1$,
$\bar n_{\rm th}'$ can be arbitrarily small. However, in this case, the classical analysis breaks down
and the quantum fluctuations of the cavity modes will set forth a quantum limit for the minimally achievable
$\bar n_{\rm th}'$, which will be detailed in the following quantum analysis.

\noindent{\it Quantized Hamiltonian.} The classical Hamiltonian derived above can be quantized by identifying these generalized coordinate and momentum as Heisenberg operators which satisfy following commutation relations
\be
[\hat q_j, \hat p_{j'}]=i\,\delta_{jj'},~~~~~~~~(j, j'=0,1,m).
\ee
The quantized Hamiltonian is then given by
\be
\label{9}
\hat{\cal H}=\frac{1}{2}\sum_{i=m,0,1}\hbar\omega_i(\hat{q}_i^2+\hat{p}_i^2)+\hbar G_0\hat{q}_m(\hat{q}_0\hat{q}_1+\hat{p}_0\hat{p}_1)+\hat{\cal H}_{\rm ext},
\ee
where we have added $\hat{\cal H}_{\rm ext}$ to take account of the coupling between cavity modes and external continuum
optical fields due to the finite transmission of the cavity. This Hamiltonian is convenient for discussing stationary
tripartite quantum entanglement as will be shown in Sec. \ref{IV}, for those generalized coordinates $q_i$ and $p_i$ correspond to the
amplitude and phase quadratures in the quantum optics entanglement experiments.

To discuss the ground-state cooling as will be investigated in Sec. \ref{III}, it would be illuminating to introduce annihilation operators
for the two cavity modes $\hat a\equiv(\hat q_0+i\,\hat p_0)/\sqrt{2}$ and $\hat b\equiv(\hat q_1+i\,\hat p_1)/\sqrt{2}$,
such that the normally ordered quantized Hamiltonian can be rewritten as
\be
\label{10}
\hat{\cal H}=\frac{1}{2}\hbar\omega_m(\hat{q}_m^2+\hat{p}_m^2)+\hbar\omega_{0}\hat{a}
^{\dag}\hat{a}+\hbar\omega_{1}\hat{b}^{\dag}\hat{b}+\hbar G_0\hat{q}_m(\hat{a}^{\dag}\hat{b}+\hat{b}^{\dag}\hat{a})
+\hat{\cal H}_{\rm ext}.
\ee

\section{Quantum limit for three-mode cooling \label{III}}
In this section, we will start from the Hamiltonian in Eq. \eqref{10} to derive the dynamics and discuss the quantum limit for the ground-state cooling experiments using three-mode optoacoustic interactions. As we will see, due to similar mathematical structure
as in the two-mode case, the corresponding quantum limit for three-mode cooling is identical to the resolved-sideband limit derived by
Marquardt {\it et al.} \cite{Marquardt} and Wilson-Rae {\it et al.} \cite{Rae} in the two-mode case.

\noindent{\it Equations of Motion.} The dynamics of this three-mode system can be derived from the quantum Langevin equations (QLEs). In the experiments, the $\rm TEM_{00}$ mode is driven on resonance at $\omega_0$. Therefore, we choose a rotating frame at $\omega_0$, obtaining the corresponding nonlinear QLEs as:
\begin{align}
\dot{\hat{q}}_m&=\omega_m \hat{p}_m,\\
\dot{\hat{p}}_m&=-\omega_m\hat{q}_m-\gamma_m \hat{p}_m-G_0(\hat{a}^{\dag}\hat{b}+\hat{b}
^{\dag}\hat{a})+\xi_{\rm th},\\
\dot{\hat{a}}&=-\gamma_0\,\hat{a}-i\,G_0\hat{q}_m\hat{b}+\sqrt{2\gamma_0}\,\hat{a}_{\rm in},\\
\dot{\hat{b}}&=-(\gamma_1-i\,\Delta)\hat{b}-i\,G_0\hat{q}_m\hat{a}+\sqrt{2\gamma_1}\,
\hat{b}_{\rm in}.
\end{align}
Here the ${\rm TEM}_{00}$ and ${\rm TEM}_{01}$ mode gap is given by $\Delta\equiv \omega_1-\omega_0$; $G_0(\hat{a}^{\dag}\hat{b}+\hat{b}^{\dag}\hat{a})$ corresponds to
the radiation pressure which modify
the dynamics of the acoustic mode and is also responsible for the quantum limit;
we have added thermal noise $\xi_{\rm th}$ whose correlation function, in the Markovian approximation, is given by $\langle \xi_{\rm th}(t)\xi_{\rm th}(t') \rangle=2\gamma_m \bar n_{\rm th}\delta(t-t')$. In obtaining above equations, we have also used Markovian approximation for $\hat{\cal H}_{\rm ext}$, namely
\be
\hat{\cal H}_{\rm ext}= i\,\hbar(\sqrt{2\gamma_0}\,\hat{a}^{\dag}\hat a_{\rm in}+\sqrt{2\gamma_1}\,\hat{b}^{\dag}\hat b_{\rm in}-H.c.)
\ee
with $H.c.$ denoting Hermitian conjugate.

To solve the above equations, we can linearize them by replacing every Heisenberg operator with the sum of a steady part and a small perturbed part, namely $\hat{o}=\bar o+\delta \hat o(\epsilon)$ with $\epsilon\ll 1$. We treat Brownian thermal noise $\xi_{\rm th}$, the vacuum fluctuations $\sqrt{\gamma_0}\delta \hat a_{\rm in}, \sqrt{\gamma_1}\delta \hat b_{\rm in}$ and $\delta \hat q_m$ as the order of
$\epsilon$. In the experiments, the $\rm TEM_{00}$ mode is pumped externally with a large classical amplitude $\bar a_{\rm in}$ while the $\rm TEM_{01}$ mode is not with $\bar b_{\rm in}=0$. Therefore, to the zeroth order of $\epsilon$, the steady part of the cavity modes are simply given by
\be
\bar a=\sqrt{{2}/{\gamma_0}}\,\bar a_{\rm in}=\sqrt{{2I_0}/({\gamma_0\hbar\omega_0})}, \quad \bar b=-iG_0\bar a\bar q_m.
\ee
Without loss of generality, we can set $\bar q_m=0$. Therefore, $\bar b=0$ and this allows us to
eliminate the $\rm TEM_{00}$ mode from the first-order equations, which are
\begin{align}
\delta\dot{\hat{q}}_m&=\omega_m\,\delta\hat{p}_m,\label{16}\\\label{17}
\delta\dot{\hat{p}}_m&=-\omega_m\,\delta\hat{q}_m-\gamma_m \,\delta\hat{p}_m-G_0\bar a(\delta\hat{b}+\delta\hat{b}^{\dag})+\xi_{\rm th},\\
\delta\dot{\hat{b}}&=-(\gamma_1+i\,\Delta)\delta\hat{b}-i\,G_0\bar{a}\,\delta\hat{q}_m+\sqrt{2\gamma_1}\,
\delta\hat{b}_{\rm in}.\label{18}
\end{align}
Here we have chosen an appropriate phase reference such that $\bar a_{\rm in}$ is real and positive.
The above equations can be solved in the frequency domain, namely
\begin{align}
\tilde q_m(\Omega)&=-\frac{\omega_m[\tilde F_{\rm rp}(\Omega)+\tilde \xi_{\rm th}(\Omega)]}{(\Omega^2-\omega_m^2)+i\,\gamma_m\Omega},\\\label{b}
\delta \tilde b(\Omega)&=\frac{G_0\,\bar a \,\delta \tilde q_m(\Omega)+i\sqrt{2\gamma_1} \delta \tilde b_{\rm in}(\Omega)}{(\Omega-\Delta)+i\,\gamma_1},
\end{align}
where the radiation pressure
\be
\label{BA}
\tilde F_{\rm rp}(\Omega)=\frac{2G_0^2 \,\bar a^2 \,\Delta\,\delta \tilde q_m(\Omega) -2G_0\bar a\sqrt{\gamma_1}[(\gamma_1-i\Omega)\delta \tilde q_2(\Omega)-\Delta\,\delta \tilde p_2(\Omega)] }{[(\Omega-\Delta)+i\,\gamma_1][(\Omega+\Delta)+i\,\gamma_1]}
\ee
with amplitude and phase quadratures $\delta\tilde q_2(\Omega)=[\delta \tilde b(\Omega)+\delta\tilde b^{\dag}(-\Omega)]/\sqrt{2}$ and
$\delta\tilde p_2(\Omega)=[\delta \tilde b(\Omega)-\delta\tilde b^{\dag}(-\Omega)]/(\sqrt{2}i)$.
In the expression of $\tilde F_{\rm rp}$, the part proportional to $\delta \tilde q_m$ is the optical spring effect. For
a high quality-factor oscillator with $\omega_m\gg\gamma_m$, the decay rate $\gamma_m$ and the eigenfrequency $\omega_m$ of
the acoustic mode will be modified to new effective $\gamma_m'$ and $\omega_m'$, as given by
\begin{align}
\gamma_m'&=\gamma_m+\frac{4G_0^2\,\bar a^2\,\Delta\,\omega_m\gamma_1}{[(\omega_m-\Delta)^2+\gamma_1^2][(\omega_m+\Delta)^2+\gamma_1^2]},\\
\omega_m'&=\omega_m+\frac{G_0^2\bar a^2\Delta(\omega_m^2-\Delta^2-\gamma_1^2)}{[(\omega_m-\Delta)^2+\gamma_1^2][(\omega_m+\Delta)^2+\gamma_1^2]}.
\end{align}
In our case, the ${\rm TEM}_{00}$ and ${\rm TEM}_{01}$ mode gap is $\Delta=\omega_1-\omega_0=\omega_m$. For the
resolved-sideband with $\gamma_1\ll\omega_m$, we obtain
\begin{align}
\gamma_m'\approx\gamma_m+\frac{G_0^2\bar a^2}{\gamma_1};\quad \omega_m'\approx \omega_m-\frac{G_0^2\bar a^2}{4\omega_m}.
\end{align}
If we define the parametric gain as ${\cal R}=(\gamma_m-\gamma_m')/\gamma_m$, then in this case
\be
{\cal R}=-\frac{G_0^2\bar a^2}{\gamma_1\gamma_m}=-\frac{2\Lambda I_0 \omega_1}{m\omega_m L^2\gamma_0\gamma_1\gamma_m}.
\ee
This is identical to Eq. \eqref{R} in the negative-gain regime, which was obtained from classical analysis by Braginsky {\it et al.} \cite{braginsky}. However, different from Eq. \eqref{nb}, the resulting thermal occupation number of the acoustic mode is
given by
\be
\bar n_{\rm th}'=\frac{\bar n_{\rm th}\gamma_m}{\gamma_m'}+\bar{n}_{\rm quant}=\frac{\bar n_{\rm th}}{1-{\cal R}}+\bar{n}_{\rm quant}
\ee
where the extra term $\bar{n}_{\rm quant}$ originates from the vacuum fluctuations in $F_{\rm rp}$, i.e. terms proportional to $\delta\tilde p_2$ and $\delta\tilde q_2$. Since in the case of large $\cal R$ or equivalently strong optoacoustic coupling, $\bar n_{\rm th}'\approx \bar{n}_{\rm quant}$ and the acoustic mode will finally reach a thermal equilibrium with the cavity modes. The lowest achievable thermal occupation number $\bar n_{\rm quant}$ will be determined by this optical heat bath (cavity mode $+$ external continuum mode).

To derive this quantum limit $\bar{n}_{\rm quant}$, we will apply the Fluctuation-Dissipation-Theorem (FDT). Specifically, given any two quantities $\hat A(t)$ and $\hat B(t)$ which linearly depend on field strength, we can define the forward correlation function
\be
C_{\hat A \hat B}(t-t')\equiv \langle \hat A(t) \hat B(t')\rangle ~~~~~ (t>t'),
\ee
where $\langle\;\rangle$ denotes the ensemble average. According to the FDT, we have
\be
\label{20}
\frac{S_{\hat A\hat B}(\Omega)+S_{\hat A\hat B}(-\Omega)}{S_{\hat A\hat B}(\Omega)-S_{\hat A\hat B}(-\Omega)}=
\frac{e^{\beta\,\hbar\,\Omega}+1}{e^{\beta\,\hbar\,\Omega}-1}=2\bar{n}_{\rm eff}(\Omega)+1,
\ee
or equivalently,
\be
\label{21}
\bar{n}_{\rm eff}(\Omega)=\frac{S_{\hat A\hat B}(-\Omega)}{S_{\hat A\hat B}(\Omega)-S_{\hat A\hat B}(-\Omega)}.
\ee
where $S_{\hat A\hat B}(\Omega)$ is the power spectral density (Fourier transform of $C_{\hat A \hat B}$) and $\beta=1/(k_B T_{\rm eff})$ and the effective thermal occupation number $\bar n_{\rm eff}\equiv1/(e^{\beta\,\hbar\,\Omega}-1)$. In our case, we can simply substitute $A, B$ with the amplitude of the $\rm TEM_{01}$ mode $\delta \hat b$ by fixing $\hat q_m=0$. From Eq. \eqref{b} and
using the fact that for vacuum fluctuation $\langle\delta\tilde b_{\rm in}(\Omega)\delta\tilde b_{\rm in}^{\dag}(\Omega')\rangle=2\pi \delta(\Omega-\Omega')$, we obtain
\be
\label{22}
S_{\delta\hat b \,\delta\hat b}(\Omega)=\frac{2\gamma_1}{(\Omega-\Delta)^2+\gamma_1^2}.
\ee
Since the acoustic mode have a very high intrinsic quality factor ($\omega_m\gg\gamma_m$),
the energy transfer between the cavity modes and the acoustic mode only happens around $\omega_m$.
Therefore, from Eq. \eqref{21} and Eq. \eqref{22}, the final quantum limit is given by
\be
\bar{n}_{\rm quant}\approx\bar{n}_{\rm eff}(\omega_m)=\left(\frac{\gamma_1}{2\omega_m}\right)^2,
\ee
where we have used the fact that for the resonant case, $\Delta=\omega_1-\omega_0=\omega_m$.
To achieve the quantum ground state, i.e. $\bar n_{\rm quant}\sim 0$, we require $\omega_m\gg\gamma_1$ and this is simply the resolved-sideband limit obtained in the pioneering works of Marquardt {\it et al.} \cite{Marquardt} and Wilson-Rae {\it et al.} \cite{Rae}.

\begin{figure}
\includegraphics[width=0.5\textwidth, bb=77 290 522 480,clip]{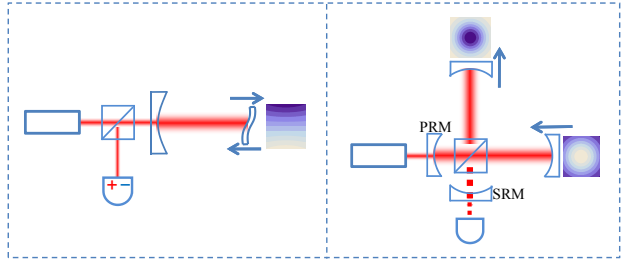}\caption{Equivalent mapping from three-mode system to an power and signal-recycled interferometer. The $\rm TEM_{00}$ and $\rm TEM_{01}$ mode can be viewed as the common and differential optical modes in the interferometer respectively. The torsional acoustic mode is equivalent to the differential motion of two end mirrors in the interferometer. By adjusting the positions of power-recycling mirror (PRM) and signal-recycling mirror (SRM), we can make the carrier and sideband modes simultaneously resonate inside the cavity, the same as the three-mode scheme.}
\label{mapping}
\end{figure}
The reason why the quantum limit for three-mode cooling is identical to the two-mode case can be readily understood from the
fact that the ${\rm TEM}_{00}$ mode is eliminated from the optoacoustic dynamics as shown explicitly in Eq. \eqref{16}\eqref{17}
\eqref{18} and we essentially obtain an effective two-mode system.
As suggested by Chen \cite{chen}, this equivalence can be more obvious by mapping this three-mode system into a power and signal-recycled laser interferometer,
as shown in Fig. \ref{mapping}. The $\rm TEM_{00}$ and $\rm TEM_{01}$ modes can be viewed as the common and differential modes in the interferometer respectively.
The torsional mode corresponds to the differential motion of the end mirrors and $\Delta$ is equivalent to the detuning of the signal recycling cavity.
In the power and signal-recycled interferometer,
even though there is no high-order transverse optical mode involved, the two degrees of freedom of the power recycling mirror and
the signal recycling mirror enable simultaneous resonances of the carrier and sideband modes, which is achieved naturally with the
three-mode optoacoustic scheme.

The above discussion shows that mathematically two-mode interactions and three-mode interactions are very similar. However,
it is very important to emphasis that from an experimental point of view, there is an important difference. Specifically, the steady-state amplitude $\bar a$ in the radiation pressure $F_{\rm rp}$ is amplified by the optical resonance while for the two-mode case this amplitude is highly suppressed due to large detuning. In other words, in order to achieve the same optoacoustic coupling strength experimentally, the input optical power in the two-mode scheme needs to be $1+(\Delta/\gamma_0)^2$ times larger than the three-mode scheme. This is a large factor in the resolved-sideband regime with $\Delta\gg\gamma_0$ (the optimal $\Delta=\omega_m$). Besides, in three-mode interactions, the condition $\Delta=\omega_m$ also naturally optimizes the energy transfer from the acoustic mode to the cavity mode \cite{Marquardt, Rae, Genes}. Therefore, three-mode scheme greatly enhances the optoacoustic coupling and is able to achieve resolved-sideband limit without compromising the intra-cavity optical power. As mentioned in Ref. \cite{zhao2}, the amplitude and laser phase noise can also be reduced significantly with three-mode scheme, simply due to the filtering of the cavity resonance.

To motivate future cooling experiments with three-mode interaction, now we present an experimentally achievable specification for the quantum ground state cooling of a milligram scale mechanical oscillator. We choose that the mass of the mechanical oscillator $m=0.1\,{\rm mg}$; the length of the cavity $L=2\,{\rm cm}$; the acoustic-mode frequency $\omega_m/2\pi=10^6\,{\rm Hz}$, the acoustic-mode quality factor $Q_m\equiv\omega_m/\gamma_m=10^7$; the optical finesse ${\cal F}=10^4$. Given an input optical power of the $\rm TEM_{00}$ mode $I_0=50\,{\rm mW}$ and the environmental temperature $T=4\,{\rm K}$, the corresponding effective thermal occupation number of the mechanical oscillator $\sim 0.5$.

\section{Stationary tripartite optoacoustic quantum entanglement \label{IV}}
As shown in the works of Vitali {\it et al.} \cite{Vitali} and Paternostro {\it et al.} \cite{Paternostro}, optoacoustic
interaction provides an very efficient way of generating stationary quantum entanglements among cavity modes and the acoustic
mode. Once experimentally realized, it will have significant impacts on future quantum communications.
Following their formulism, we will investigate the stationary tripartite quantum entanglement in the three-mode optoacoustic system by first analyzing the dynamics and then evaluating the entanglement measure --- logarithmic negativity $E_{\cal N}$ defined in Ref. \cite{Vidal, Adesso}.

Starting from Hamiltonian in Eq. \eqref{9}, the corresponding nonlinear QLEs in the rotating frame at the laser frequency $\omega_L$ can be written as
\begin{align}
\dot{\hat{q}}_m&=\omega_m \hat{p}_m,\\\label{36}
\dot{\hat{p}}_m&=-\omega_m\hat{q}_m-\gamma_m \hat{p}_m-G_0(\hat{q}_0\hat{q}_1+\hat{p}_0\hat{p}_1)+\xi_{\rm th},\\
\dot{\hat q}_0&= -\gamma_0\hat{q}_0+\Delta_0 \hat p_0+G_0\hat q_m \hat p_1+\sqrt{2\gamma_0}\hat q_0^{\rm in},\\
\dot{\hat{p}}_0&=-\gamma_0\hat p_0-\Delta_0\hat q_0-G_0\hat{q}_m\hat{q}_1+\sqrt{2\gamma_0}\,\hat{p}_0^{\rm in},\\
\dot{\hat q}_1&= -\gamma_1\hat q_1+\Delta_1 \hat p_1+G_0\hat q_m \hat p_0+\sqrt{2\gamma_1}\hat q_1^{\rm in},\\
\dot{\hat{p}}_1&=-\gamma_1\hat p_1-\Delta_1 \hat q_1-G_0\hat{q}_m\hat{q}_0+\sqrt{2\gamma_1}\,\hat{p}_1^{\rm in},
\end{align}
where $\Delta_{0}=\omega_0-\omega_L$ and $\Delta_1=\omega_{1}-\omega_L$.
Slightly different from the cooling experiments, here we need to externally drive both
the $\rm TEM_{\rm 00}$ and $\rm TEM_{01}$ mode simultaneously to create tripartite quantum entanglement. We choose an appropriate phase reference such that the classical amplitude $\bar{p}_{i}=0$ and $\bar{q}_{i}\neq 0~~(i=0,1)$ which is related to the input optical power $I_i$ by $ \bar{q}_{i}=\sqrt{2 I_{i}/(\hbar\omega_i\gamma_i)}$. Similar to the previous case, we can linearize above equations as
\be\label{41}
\dot {\hat{\bf x}}^{\bf T}={\mathbf M}\, \hat{\bf x}^{\bf T}+\hat{\mathbf  n}^{\bf T},
\ee
with $\bf T$ denoting a transpose transformation and
\begin{align}
\hat{\mathbf x}^{\bf T}&\equiv(\begin{array}{cccccc}\delta \hat q_m,
&\delta \hat p_m,&\delta \hat q_0,&\delta \hat p_0,&\delta \hat q_1,&\delta \hat p_1\end{array})^{\bf T},\\
\hat{\mathbf n}^{\bf T}&\equiv(\begin{array}{cccccc}0,
&\xi_{\rm th},&\sqrt{2\gamma_0} \,\delta \hat q_0^{\rm in},&\sqrt{2\gamma_0}\, \delta \hat p_0^{\rm in},&\sqrt{2\gamma_1}
\,\delta \hat q_1^{\rm in},&\sqrt{2\gamma_1}\,\delta\hat p_1^{\rm in}\end{array})^{\bf T}
\end{align}
and matrix $\mathbf M$ is given by
\be
{\mathbf M}=\left(\begin{array}{cccccc}
           0&\omega_m&0&0&0&0\\
           -\omega_m&-\gamma_m&G_0\bar{q}_1&0&G_0\bar{q}_0&0\\
           0&0&-\gamma_0&\Delta_0&0&0\\
           G_0\bar{q}_1&0&-\Delta_0&-\gamma_0&0&0\\
           0&0&0&0&-\gamma_1&\Delta_1\\
           G_0\bar{q}_0&0&0&0&-\Delta_1&-\gamma_1
           \end{array}\right).
\ee
At first sight, the mathematical structure is identical to the one analyzed by Paternostro {\it et al.} \cite{Paternostro}.
Apart from differing in the coupling constants (here we need to consider the overlapping factor $\Lambda$), there
is another important difference: After linearization, the radiation pressure term $G_0(\hat q_0\hat q_1+\hat p_0\hat p_1)$ in Eq. \eqref{36} is proportional to $\bar q_0\delta\hat q_1+\bar q_1\delta\hat q_0$ rather than $\bar q_0\hat q_0-\bar q_1\hat q_1$ considered in Ref. \cite{Paternostro}. As we will show, similar to the case for cooling experiments, the coherent build up of both the ${\rm TEM}_{00}$ and ${\rm TEM}_{01}$ mode and optimal mode gap $\omega_1-\omega_0=\omega_m$ enhance the entanglement significantly, which make it easier to achieve experimentally.

Assuming the system is stable, i.e. all eigenvalues of $\bf M$ have negative real part, the stationary solutions to Eq. \eqref{41}
can be written down formally as
\be
\hat x_i(\infty)=\sum_j\int_0^{\infty}dt' [e^{{\mathbf M}(t-t')}]_{ij}\hat n_j(t'),
\ee
where we have neglected the initial-condition terms which decay away as the system approaches the stationary state.
We assume that all the noises are Markovian Gaussian processes and the correlation functions are \be\sigma_{ij}(t-t')\equiv D_{ij}\,\delta(t-t')\ee where $D_{ij}$ are the elements of matrix $\bf D$ and $D_{ij}={\rm Diag}[0, 2\gamma_m k_B T/(\hbar\omega_m),\gamma_0,\gamma_0,\gamma_1,\gamma_1]$.
The corresponding stationary covariance matrix among the cavity modes and the acoustic mode can then be written as
\be
{\mathbf V}(\infty)=\int_0^{\infty} dt [e^{{\mathbf M}t}]{\mathbf D}[e^{{\mathbf M}t}]^{\bf T},
\ee
and the components of $\mathbf V$ can be obtained by solving following algebra equations:
\be
\mathbf M\,\mathbf V+\mathbf V\,\mathbf M^{\bf T}=-\mathbf D.
\ee

For this tripartite continuous-variable system (one acoustic mode $+$ two cavity modes), one necessary and sufficient condition for
separability is the positivity of partially transposed covariance matrix \cite{Pere, Simon, Serafini}. In our case, partial transpose
is equivalent to time reversal and can be realized by reverting the momentum of the acoustic mode form $\hat p_m$ to $-\hat p_m$, namely
\be
{\bf V}_{\rm pt}={\bf V}|_{\hat p_m\rightarrow-\hat p_m}.
\ee
By evaluating the positivity of the eigenvalue of ${\bf V}_{\rm pt}$, we can directly determine whether entanglement exists or not. To reveal the richness of the entanglement structure, we will not directly analyze the positivity of $\bf V_{\rm pt}$ for the entire system, but rather following Ref. \cite{Paternostro}, we look at the entanglement between any bipartite subsystem
using the logarithmic negativity $E_{\cal N}$. Given the $4\times 4$ covariance matrix ${\bf V}_{\rm sub}$ for any bipartite subsystem,
\be
{\bf V}_{\rm sub}=\left[\begin{array}{cc}{\bf A}_{2\times 2}&{\bf C}_{2\times 2}\\
{\bf C}_{2\times 2}^{\bf T}&{\bf B}_{2\times 2}\end{array}\right],
\ee
the logarithmic negativity $E_{\cal N}$ is defined by \cite{Adesso, Pere}
\be
E_{\cal N}=\max[0,-\ln 2\sigma_-]
\ee
with $\sigma_-\equiv \sqrt{\Sigma-\sqrt{\Sigma^2-4\det {\bf V}_{\rm sub}}}/\sqrt{2}$ and
$\Sigma\equiv \det {\bf A}+\det {\bf B}-2\det {\bf C}$.

\begin{figure*}
\includegraphics[width=0.3\textwidth,bb=0 0 309 315,clip]{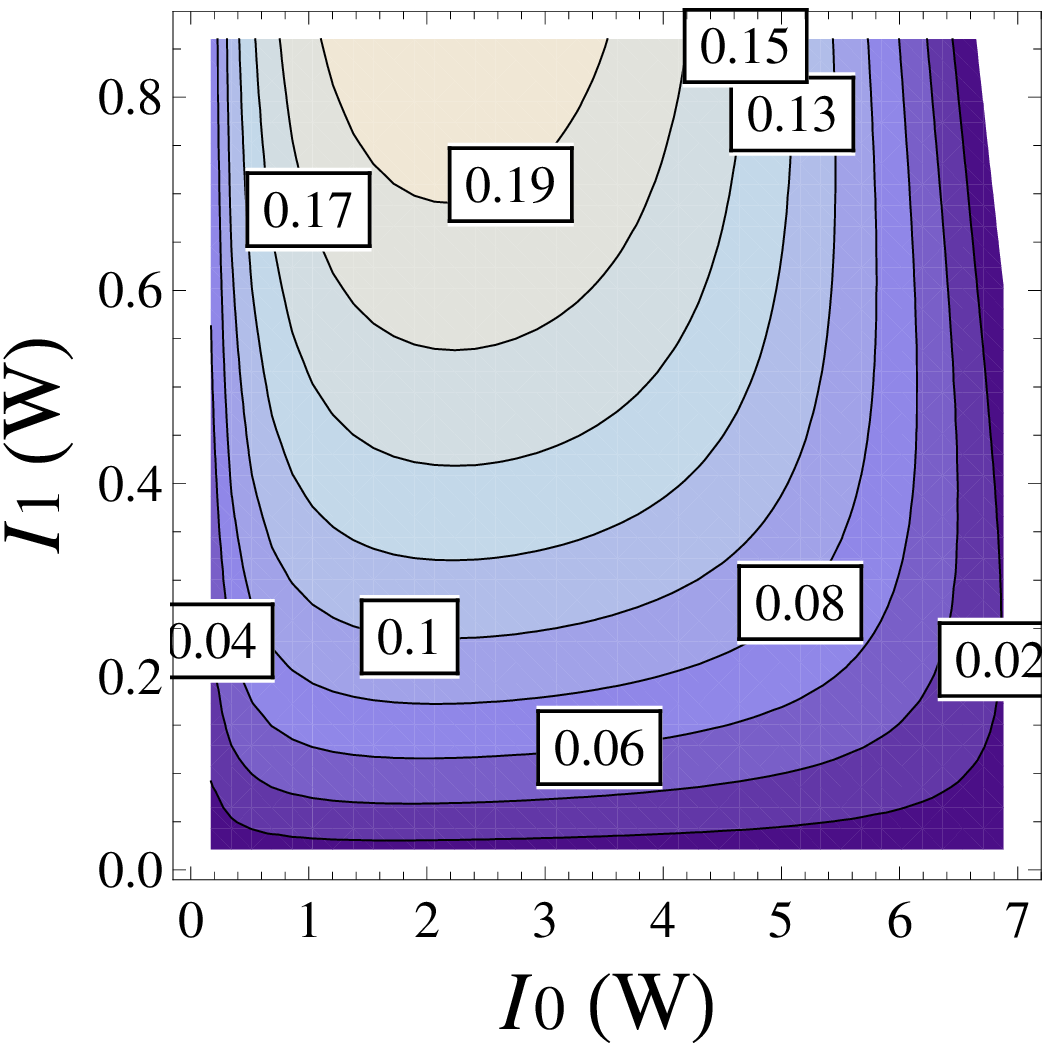}
\includegraphics[width=0.3\textwidth,bb=0 0 309 315,clip]{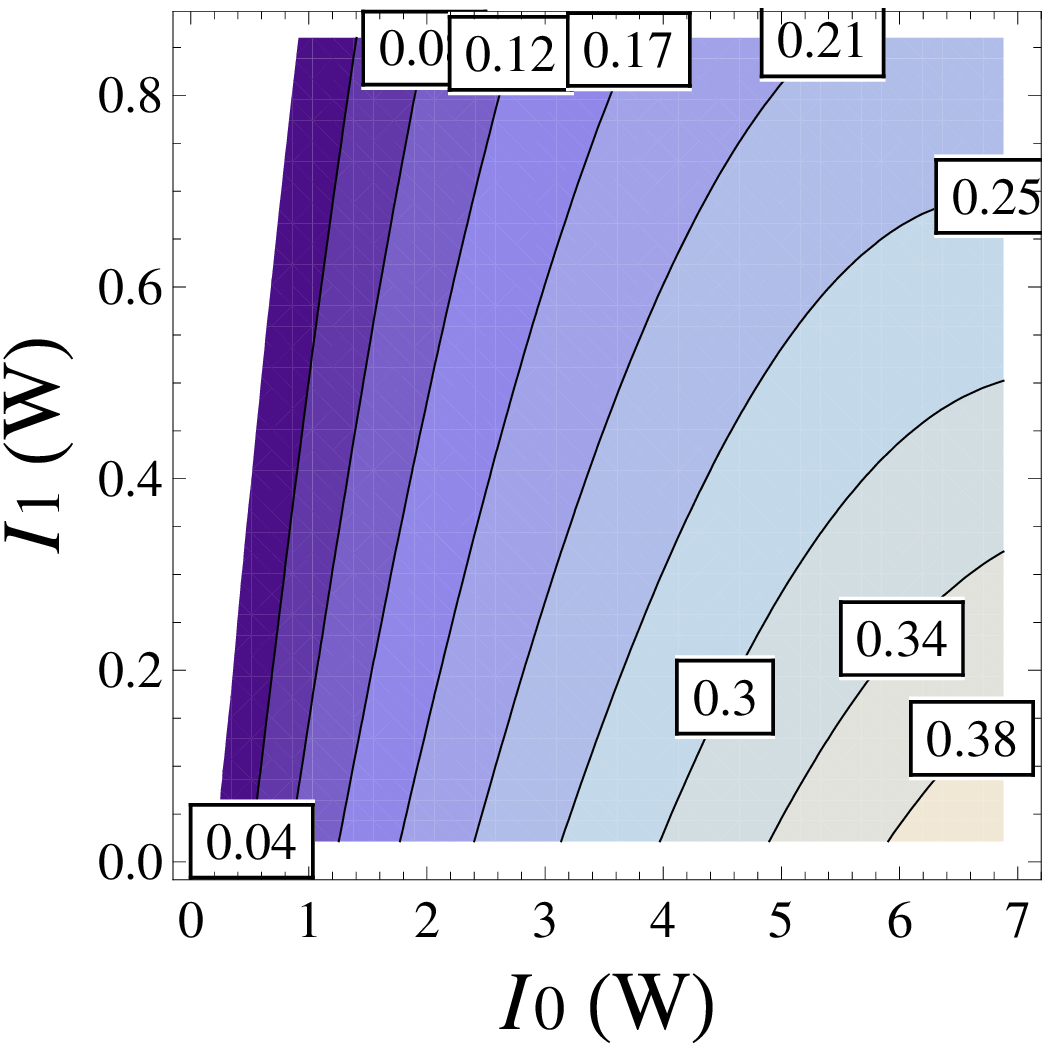}
\includegraphics[width=0.3\textwidth,bb=0 0 309 315,clip]{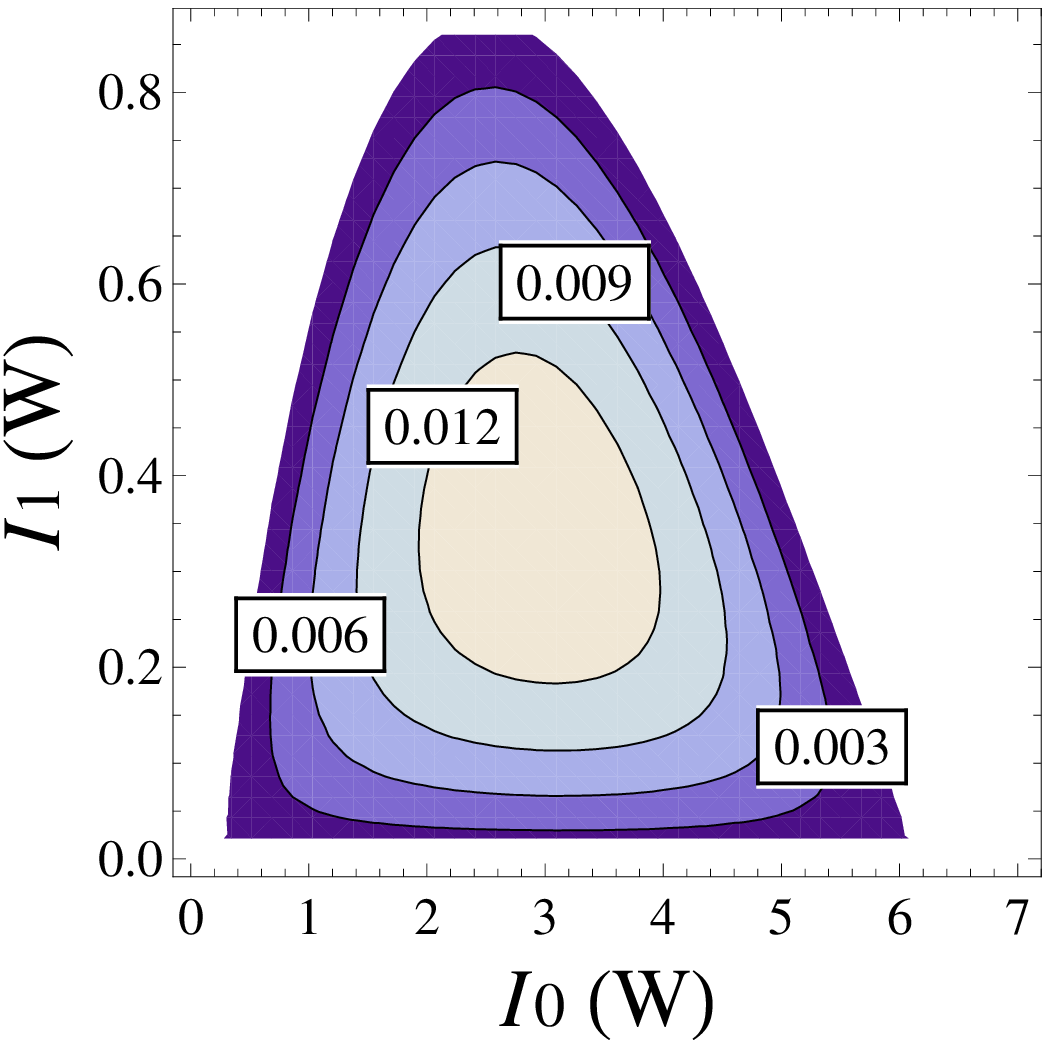}\caption{Logrithmic negativity ${E}_{\cal N}$
as a function of the input optical powers of both modes. Other specifications are identical to those for cooling experiments given in the previous section.
The left panel shows ${ E}^{\rm 0m}_{\cal N}$ for the entanglement between the $\rm TEM_{00}$ mode and the acoustic mode; The middle panel presents ${E}^{\rm 1m}_{\cal N}$ for the $\rm TEM_{01}$ mode and the acoustic mode; The right panel shows ${ E}^{01}_{\cal N}$ for the $\rm TEM_{00}$ mode and the $\rm TEM_{01}$ mode.  \label{EN}}
\end{figure*}

\begin{figure}
\includegraphics[width=0.45\textwidth, bb=0 0 472 300,clip]{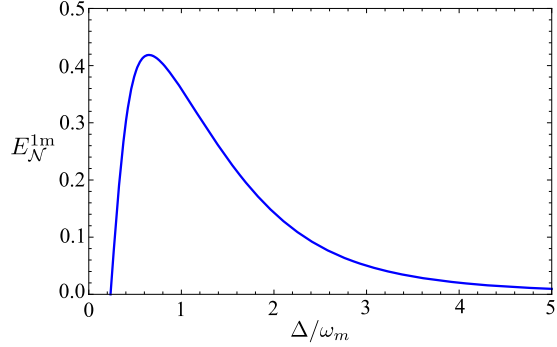}
\caption{Logarithmic negativity ${E}^{\rm 1m}_{\cal N}$ as a function of cavity modes gap $\Delta\equiv\omega_1-\omega_0$. As we can see,
the condition $\Delta\approx \omega_m$, which optimizes the cooling, also maximizes the entanglement between the $\rm TEM_{01}$ mode and the acoustic mode. Here
we have assumed $I_0=4.5\,{\rm W}$ (Higher $I_0$ will make the system unstable for small $\Delta$) and $I_1=0.\,{\rm W}$. Since it
can be viewed as an effective two-mode system in this case with $I_1=0$, it simply recovers the results given by Vitali {\it et al.} \cite{Vitali}. \label{END}}
\end{figure}

For numerical estimations, we will use the same specification as given in the previous section for the cooling experiments. We will
focus on the situation relevant to the experiments with $\omega_1-\omega_0=\omega_m$ and the $\rm TEM_{00}$ mode driven on resonance ($\Delta_0=0,\, \Delta_1=\omega_m$).  In Fig. \ref{EN}, we show the resulting $E_{\cal N}$ as a function of the input optical powers of both optical modes. Given the specifications, the entanglement strength between each optical mode and the acoustic mode becomes stronger as the optical power of their counterpart increases (until the system becomes unstable). This is understandable, because we have $\hat q_m(\hat q_0\hat q_1+\hat q_0\hat q_1)$ type of interaction, and the coupling strength between the $\rm TEM_{00}$ mode and the acoustic mode directly depends on the classical amplitude of the $\rm TEM_{01}$ and vice versa.
For the entanglement between two optical modes, it reaches maximum when both optical modes have medium power. This can be attributable to the fact that the entanglement between the two optical mode is mediated by the acoustic mode, and both ${ E}^{\rm 0m}_{\cal N}$ and ${ E}^{\rm 1m}_{\cal N}$ should be large to give a reasonable ${ E}^{01}_{\cal N}$.
Besides, as shown explicitly in Fig. \ref{END}, the condition $\omega_1-\omega_0=\omega_m$ will naturally optimizes the entanglement between the $\rm TEM_{01}$ mode and the acoustic mode. This is because the Lorentzian profiles of the $\rm TEM_{01}$ mode and the acoustic have the largest overlap when $\Delta=\omega_m$. In this case, both the $\rm TEM_{01}$ mode and the acoustic mode are driven by the same vacuum field, which gives the maximal entanglement. Therefore, the optimal condition for the cooling experiment will simultaneously optimize the entanglement strength, as also been observed by Genes {\it et al.} \cite{Genes2}.

\begin{figure}
\includegraphics[width=0.45\textwidth, bb=0 0 472 300,clip]{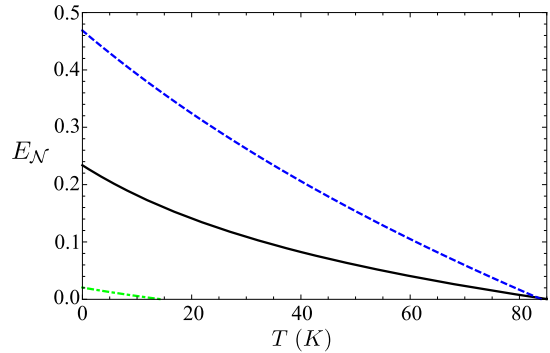}
\caption{Logarithmic negativity as a function of temperature. The solid curve stands for ${E}^{\rm 0m}_{\cal N}$, the
dashed curve for ${E}^{\rm 0m}_{\cal N}$ and dash-dot curve for ${E}^{01}_{\cal N}$. We have chosen the optimal
parameters for each curve. \label{ENT}}
\end{figure}

To illustrate the robustness of this tripartite entanglement, we show the dependence of $E_{\cal N}$ on the environmental temperature in Fig. \ref{ENT}. The entanglement between the optical modes and the acoustic mode is very robust and it persists even when the temperature goes up to $80$ K.
Although the entanglement between the two optical modes is relatively weak, yet it changes slower as the temperature increases, and it vanishes when the temperature becomes higher than $15$\,K. The robustness of the optoacoustic entanglement was also shown previously by Vitali {\it et al.} \cite{Vitali}. This is attributable to the strong optoacoustic coupling which suppresses the thermal decoherence of the acoustic mode.  With both the $\rm TEM_{00}$ mode and  the $\rm TEM_{01}$ mode on resonance, we can obtain much higher intra-cavity power compared with the equivalent detuned two-mode system. Given moderate input optical power, this allows us to achieve stronger entanglement between the optical modes and the acoustic mode of a massive mechanical oscillator ($\sim {\rm mg}$). Of course, this robustness of entanglement is conditional on the fact that the mirrors of the cavity can sustain a high optical power $\sim 10^4 \,W$. If the beam size is
of the order of mm, this corresponds to a power density of around $10^6\,{\rm W/cm}^2$, which is achievable with the present technology \cite{Reitze}.

To verify this tripartite entanglement experimentally, we can apply the same protocol as proposed in Ref. \cite{Vitali,Paternostro,Laurat}. Specifically, through measuring the outgoing field, we can build up statistics and construct the covariance matrix ${\bf V}_{\rm exp}$ of this tripartite system based on the measurement results and then analyze whether the partially transposed covariance matrix ${\bf V}_{\rm exp}^{\rm pt}$ fails to be positive definite. If ${\bf V}_{\rm exp}^{\rm pt}$ has
a negative eigenvalue, this will give an unambiguous signature for quantum entanglement, because any classical correlation
always gives a positive definite ${\bf V}_{\rm exp}^{\rm pt}$. Besides, we can also use ${\bf V}_{\rm exp}$ to evaluate the logarithmic negativity $E_{\cal N}$ of any bipartite subsystem to determine whether entanglement exists or not in a given
subsystem. Since the tripartite entanglement is stationary, this means that the optoacoustic interactions protect the quantum entanglement from the thermal decoherence which is a significant issue in non-stationary quantum entanglements. In principle, we can make a sufficiently long integration of the output signal such that the shot noise is
negligibly small and ${\bf V}_{\rm exp}$ would be a direct verification of what we have obtained theoretically.

\section{Conclusion \label{V}}
We have analyzed the three-mode optoacoustic parametric interactions in the quantum picture.  We have
derived the quantum limit for the cooling experiments with three-mode interactions based upon the Fluctuation-Dissipation-Theorem. Besides, we have shown the existence of the tripartite quantum entanglements in this system. The simultaneous resonances of
the carrier and sideband modes in the three-mode system allows more efficient acoustic-mode cooling and more robust optoacoustic entanglement than the two-mode system. This work provides the theoretical basis for the feasibility of realizing both ground-state cooling and stationary optoacoustic quantum entanglements using three-mode optoacoustic parametric interactions in small-scale
table top experiments.

\section{Acknowledgements}
We thank Prof. Y. Chen for stimulating discussions. This research has been supported by the Australian Research Council
and the Department of Education, Science and Training and by the U.S. National Science Foundation. We thank the LIGO
Scientific Collaboration International Advisory Committee of the Gingin High Optical Power Facility for their supports.
H. Miao thanks Prof. Y. Chen for the invitation to visit the Albert-Einstein-Institut and Caltech. The visits were supported by Alexander von
Humboldt Foundation's Sofja Kovalevskaja Programme and NSF grants PHY-0653653 and PHY-0601459, as well as the David and Barbara Groce startup fund at Caltech. H. Miao would also like to thank Prof. A. Heidmann, Prof. P. F. Cohandon and Dr. C. Molinelli for hosting his visit to Laboratoire Kastler Brossel at Paris and the fruitful discussions during the visit.

\end{document}